\documentclass{article}
\usepackage{graphics}
\usepackage{latexsym}
\newcommand{\be}{\begin{equation}}
\newcommand{\ee}{\end{equation}}
\newcommand{\Su}{S_F^\alpha}
\newcommand{\bw}{\mathbf{w}}

\newcommand{\tf}{\tilde{f}}
\newcommand{\R}{\mathbf{R}}
\newtheorem{definition}{Definition}
\title{Random Walk and Broad Distributions on Fractal Curves}
\author{Seema E Satin \and A.D.Gangal}
\begin{document}
\maketitle
\begin{abstract}
In this paper we analyse random walk on a fractal structure, specifically 
fractal curves, using the recently develped calculus for fractal curves. We 
consider only unbiased random walk on the fractal stucture and find out the
 corresponding probability distribution which is gaussian like in nature, but
 shows deviation from the standard behaviour. Moments are calculated in terms
 of Euclidean distance for a von Koch curve. We also analyse Levy distribution
on the same fractal structure, where the dimension of the fractal curve shows
significant contribution to the distrubution law by modyfying the nature of
moments. The appendix gives a short note on Fourier transform on fractal
curves.
\end{abstract}
\section{Introduction}
Random walks in one dimesion is a well established topic 
\cite{Chandra,van,Grimmet,Reif}, with several applications in physics including
transport phenomena and diffusive behaviour.
 The discrete simple random walk in 1-dimension is described by a particle
 taking  
steps, each towards the right or left with a specified probability. For unbiased
 walks the
 probability is same for right or left step \cite{Grimmet}. In the continuum  
limit the  number of steps $N$ goes to $\infty$, the step size $\delta$ goes 
to zero, such that $N\delta$ remains fixed. The random variable is 
rescaled to give the probability distribution \cite{van,Grimmet}, showing the 
diffusive behaviour. 

While studying transport properties on disordered systems, one encounters 
subdiffusive behaviour \cite{Bouchaud}. Some  physical examples are, NMR 
diffusometry on percolation stuctures
\cite{Muller}, motion of a bead in polymer network \cite{Amblard} etc.
In general  the diffusion process can be expected to depend on both the 
geometry of the medium and the process. While various aspects of dependence on 
processes (e.g. Gaussian, Levy etc) and ordinary geometries are well studied, 
the dependence on fractal geometries of media is relatively unexplored.

 It is expected that on fractals, subdiffusion can be due to the geometry of 
the
 underlying fractal stucture on which the process takes place.
In this chapter, various well established approaches in random walk problems
are extended to fractal geometry. We study such a behaviour by
 considering a random walk on a fractal curve
. In particular, we  present the analytical and numerical results for walks on a
 von Koch 
curve in $\R^2$.
Random walk on such a curve is performed by taking a step either forward or
backward along the curve itself with equal probability if the walk is unbiased.
Here we consider unbiased walk only.

Such a random walk in the continuum limit leads to study of processes like
 diffusion on the Fractal curve $F$. Hence one can study the diffusive
behaviour of particles, performing random motion on the fractal curve, by using
this formulation. Appropriate and exact expressions can be obtianed,
for a class of walks, to describe physical processes by using this model. The simplest case is that of 
obtaining a probability distribution, which is same as the solution obtained
by solving a diffusion equation on a fractal curve as in \cite{Seema}. The 
study of a
discrete random walk on a fractal curve $F$ is another approach towards
 studying such transport properties on these kind of structures.

\begin{figure}[h!]
\resizebox{\columnwidth}{!}{\includegraphics{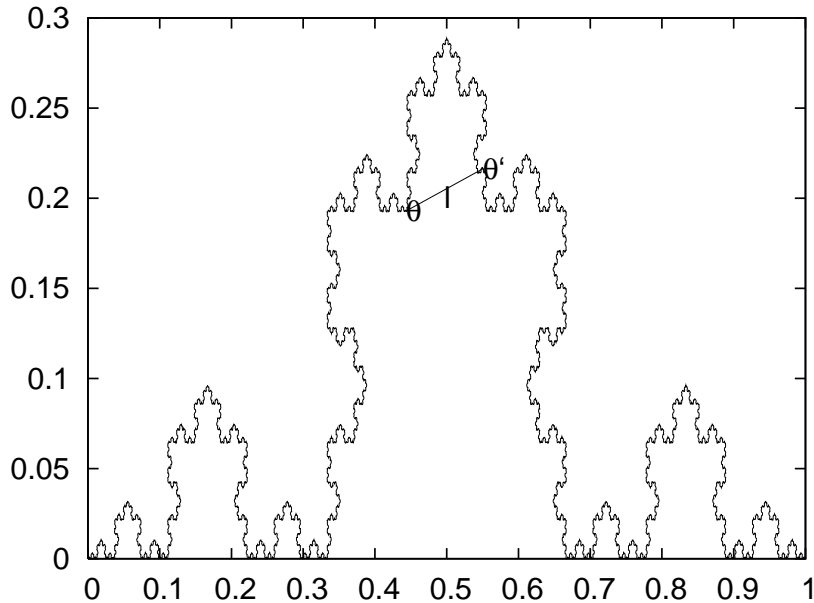}}
\caption{ Example of a fractal curve. $\theta$ and $\theta'$ denote any two
points on the curve, and $l$ gives the Euclidean distance between these two
 points }
\label{fig:labelling}
\end{figure}
There are several approaches to study anomalous  diffusion. Some commonly used
frameworks are fractional Brownian  motion \cite{Mandelbrot2}, the continuous 
time random walk \cite{ben1}, fractional diffusion equations 
\cite{Schneider,Metzler1},
 generalized Langevin equation and Fokker-Planck equations 
\cite{Fog,Fog2,Zaslav,Kiran} etc. In most of these approaches  
the assumptions for validity of simplest form of Central
 limit theorem are not satisfied \cite{Bouchaud}.

In the present case, the Central limit theorem is not violated. The 
departure from Gaussian distribution or, the normal diffusive behaviour, is
explicitly due to the underlying fractal geometry of space and not due to 
correlations or long tailed statistics.  

Here, we consider a random walk on a fractal curve(von-Koch like), in 
$\mathbf{R^n}$. The calculus on such a curve is developed in \cite{Seema}
, where the 
curve $F$ has fractal dimension $\alpha$. $\theta$ denotes a point on the 
curve and $J(\theta) \equiv S_F^\alpha(u)$ denoted the mass function, which 
is the mass of the fractal covered upto point $\theta$, $u$ is the parameter 
of the curve. A short review of the Calculus is given in Appendix 2.

Random walk on such a curve is performed by taking a step either forward or
backward along the curve itself with equal probability if the walk is unbiased.
In this paper we consider unbiased walk only.

In various sections of this paper we find analogues of simplest aspects of
random walk problems in 1-D extended to fractal curves.

\section{Discrete Random Walk on a fractal curve} \label{sec:discreet}
We consider the special class of random walks on a fractal curve $F$, where 
 the walker covers a fixed mass $\Delta$ in each step. If the fractal curve,
$F$ reduces to a real line, then the random walk on it reduces to an
ordinary simple random walk as analysed in \cite{rudnick}.

Let $C(N,\theta,\theta')$ be the number of walks that start at a point $\theta$
on the fractal curve $F$ with $\gamma$-dimension, $\alpha$ ( note 
that $\alpha \geq 1$ in general, particularly 
here $1 <\alpha < 2$ ),
and end at another point $\theta'$ on the same fractal curve, at the $N^{th}$
step. Then
\be \label{eq:recursion}
C(N;\theta,\theta') = C(N-1;\theta, J^{-1}(J(\theta') - \Delta)) + C(N-1;\theta
, J^{-1}(J(\theta') + \Delta))
\ee

Equation (\ref{eq:recursion}) is the recursion relation for a random walk
on $F$.

When $N$ i.e number of steps is large, and $\Delta$ small, equation
 (\ref{eq:recursion}) can be 
approximated by a differential equation involving the $F^\alpha$-derivative. 
This is carried out in the following.

\begin{eqnarray}
C(N+1;\theta,\theta')& = &\{C(N;\theta,J^{-1}(J(\theta')-\Delta)) + C(N;\theta,
J^{-1}(J(\theta')+\Delta))- 2 C(N;\theta,\theta')\} \nonumber \\
 & & + 2 C(N,\theta,\theta') \nonumber \\
\end{eqnarray}
Taylor expanding (as given in appendix) first two terms around
 $J(\theta')$ in powers of $\Delta$, we get
\begin{eqnarray}
C(N+1;\theta,\theta') & = & (\Delta^2 (D^\alpha_{F_{\theta'}})^2
C(N;\theta,\theta') + \frac{1}{12} \Delta^4 (D^\alpha_{F_{\theta'}})^4 
C(N;\theta,\theta') + \dots) + 2C(N;\theta,\theta') \nonumber \\
& \approx & \Delta^2 (D_{F_{\theta'}}^\alpha)^2 C(N;\theta,\theta')+ 
O(\Delta^4) + 2 C(N;\theta,\theta') \\
\end{eqnarray}
\be \label{eq:approx}
C(N+1;\theta,\theta') - 2 C(N,\theta,\theta') \approx  \Delta^2 (D_{F_{\theta'}}
^\alpha)^2 C(N;\theta,\theta')
\ee
Each step has equal probability in forward or reverse direction on the
curve.  The total number of steps being $N$, there are $2^N$ ways of 
performing an $N$ step random walk. Thus, $P(N,\theta, \theta')$,
 the probability of a such an $N$-step random walk, starting at $\theta$ and 
ending at $\theta '$ is given by
\[P(N, \theta, \theta') = \frac{1}{2^N} C(N,\theta, \theta') \]
 Hence we can replace $C(N,\theta,\theta')$ by $2^N P(N,\theta,\theta')$ and
  $C(N+1,\theta,\theta')$ by $2^{N+1} P(N+1,\theta,\theta')$.

Thus, equation (\ref{eq:approx}) leads to
\[
P(N+1,\theta,\theta') - P(N;\theta,\theta') \approx  \frac{\Delta^2}{2} 
(D_{F_{\theta'}}^\alpha)^2 P(N;\theta,\theta')
\]
We now assume $N$ to be large and $P(N,\theta,\theta')$ to be slowly varying
 function of $N$. Thus we can write the above equation in the form
\be
\frac{\partial}{\partial N } P(N;\theta,\theta') = \frac{\Delta^2}{2}
(D_{F_{\theta'}})^2 P(N,\theta,\theta')
\ee 
The solution of the above equation can easily be obtained by using conjugacy
between the $F^\alpha$-derivative and ordinary derivative as in 
\cite{Abhay}. This gives,
\be \label{eq:disprob}
P(N,\theta,\theta') = \frac{1}{\sqrt{2 \pi N}} \exp(\frac{-(J(\theta)-J(\theta')
)^2}{2 \Delta^2 N})
\ee

Now, one can  go to the continuum case, where one can set $t = N \tau$, $N$ 
being the number of 
steps in the discrete walk and $\tau$ being the duration between the 
consequtive steps.
Thus the expression for probability distribution in the continumm
case, which is of the form given by equation (\ref{eq:probdist}) in the later 
section can be obtained and used readily.
\section{Moments} \label{sec:probabdist}
We now consider the Euclidean distance $ L(\theta) \equiv L(\bw(u))= |\bw(u)|$
 upto a point $\theta =\bw(u)$ from the origin on the fractal curve $F$.

Consider the probability distribution which behaves as 
\be
P(\theta) \equiv P(\mathbf{w}(u) = \theta) \sim \exp(- S_F^\alpha(u)^2)
\mbox { where } \Su(u) = J(\theta)
\ee
with appropriate normalization.
Using this form for the probability distribution we calculate the first two
 absolute moments on the fractal curve as follows:
\be \label{eq:firstm}
< L > = \int_{C(-\infty,\infty)} L(\theta) P(\theta) d_F^\alpha \theta
\ee
and 
\be \label{eq:secondm}
< L^2 > = \int_{C(-\infty,\infty)} L^2(\theta) P(\theta) d_F^\alpha \theta
\ee

We evaluate the above two moments in terms of Euclidean distance, using the 
explicit form 
\be \label{eq:probdist}
P(w(u) = \theta) = \frac{1}{\sqrt{2 \pi A t}} \exp(-(\frac{(J(\theta) = 
S_F^\alpha(u))^2}{2 A t})) 
\ee
as obtained in chapter \cite{Seema} , which is the continuum case of 
equation \ref{eq:disprob} as $N \rightarrow \infty$ and $\Delta \rightarrow 0$
($N \Delta $ being held fixed).
\subsection*{ Heuristic Calculation of Absolute Moments}
Now, for the given gaussian like probability, equation(\ref{eq:probdist}), we
 calculate the moments in terms of Euclidean distance, for random motion on a 
von Koch curve. From the graph 
of Rise function and Euclidean distance, for a von Koch curve as given in
\cite{Abhay}, it  is clear, that 
$c_1 [L(\mathbf{w}(u))]^\alpha < S_F^\alpha(u)< c_2 [L(\mathbf{w}(u))]^\alpha$.
Thus we can approximate $S_F^\alpha (u) $ by $L^\alpha$, i.e
 
\[ S_F^\alpha (u) \sim [L(\mathbf{w}(u))]^\alpha \]
or in short hand notation
\[S_F^\alpha \sim L^\alpha \]

Thus, the first moment can be calculated  analytically by using an 
intuitive replacement for the $F^\alpha$-integrals in equations
(\ref{eq:firstm}) and (\ref{eq:secondm}). $S_F^\alpha$ is thus replaced by
$L^\alpha$ and $d_F^\alpha \theta$ is replaced by $d L^\alpha$, keeping in view
 the way $F^\alpha$-integrals are defined in \cite{Abhay} . 

Thus,
\be
<L> = 2 \frac{1}{\sqrt{2 \pi A t}}\int_{C(0,\infty)} L \exp(-\frac{(L^\alpha)^2}
{2 A t }) dL^\alpha
\ee

On substituting $L^\alpha/\sqrt{2At} =z$ we 
reduce the integral to the following
\be
<L> = \mbox{ Const.} t^{1/2 \alpha} \int_0^\infty z^{1/\alpha} \exp(-z^2) dz
\ee

which implies 

\be \label{eq:L1}
<L>  \sim t^{1/2\alpha}
\ee
 similarly we can obtain the second moment in terms of the Euclidean distance

\be
<L^2> = 2 \frac{1}{\sqrt{2 \pi A t}}\int_{C(0,\infty)} L^2 \exp(-\frac{(L^\alpha
)^2}{2 A t }) dL^\alpha
\ee
leading to the behaviour

\be \label{eq:L2}
< L^2> \sim t^{1/\alpha}
\ee

\begin{figure}[h!]
\resizebox{\columnwidth}{!}{\includegraphics{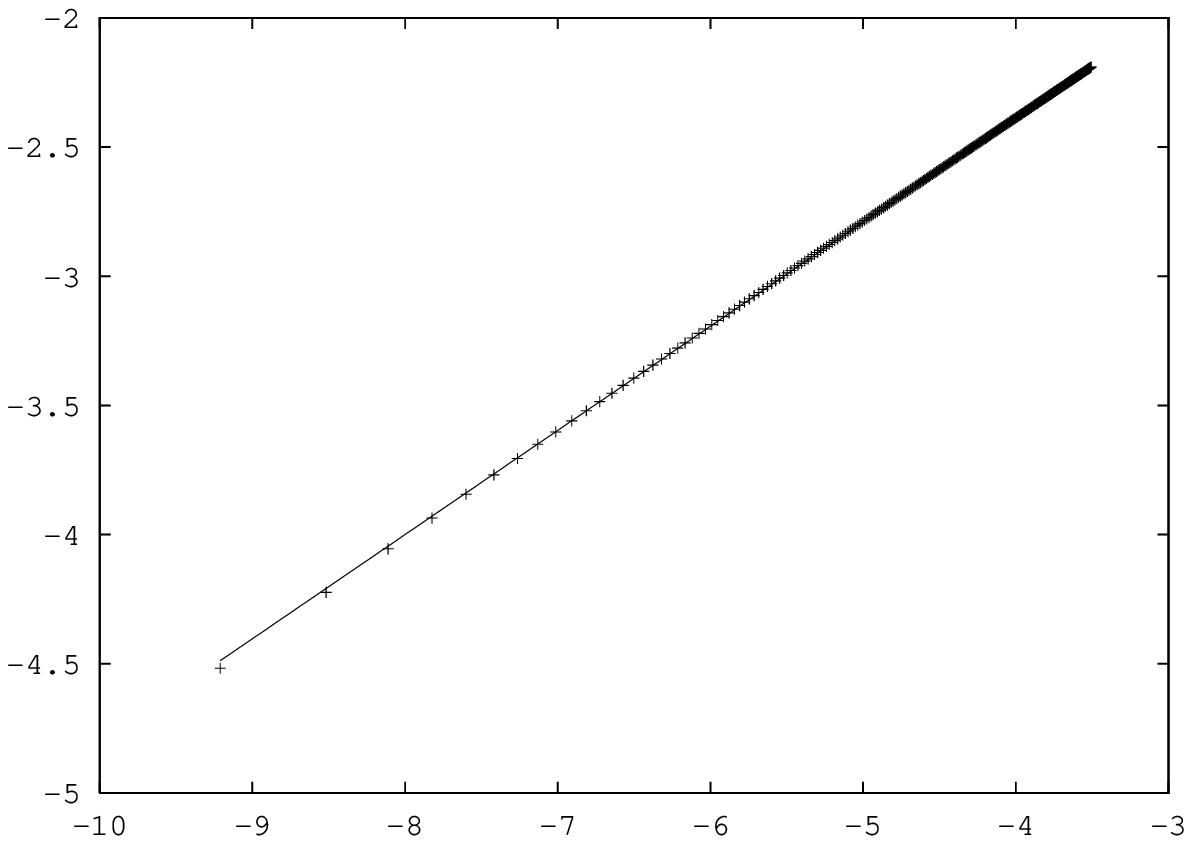}}
\caption{ Plot of $log <L>$ (Y-axis) vs. $log(t)$ (X-axis). The cross denotes
the value (calculated by performing $F^\alpha$- integration 
numerically), to which the straight  line is fitted with slope 0.403 ,
 to extract the
value of exponent in equation (\ref{eq:L1}), for a von Koch curve.}
\label{fig:firstm}
\end{figure}

\begin{figure}[h!]
\resizebox{\columnwidth}{!}{\includegraphics{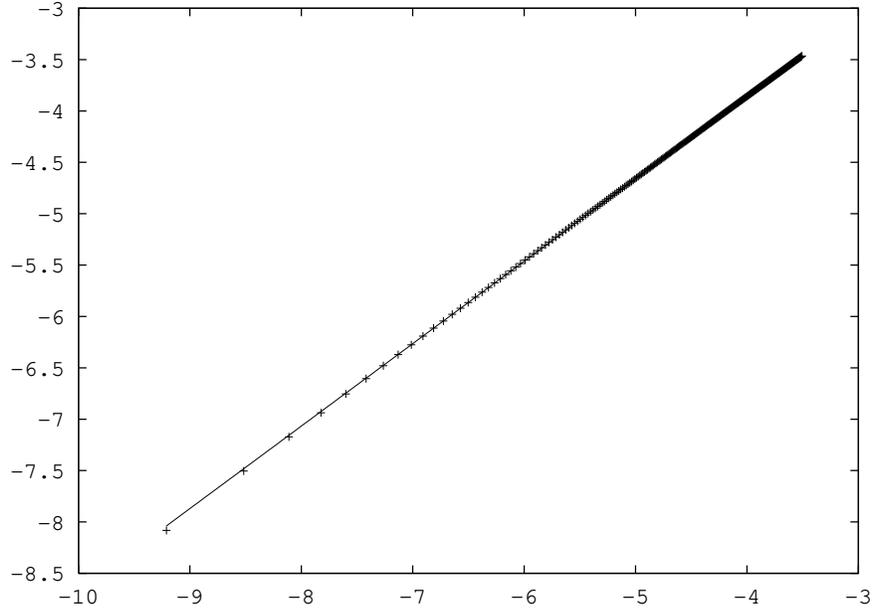}}
\caption{ Plot of $log <L^2>$ (Y-axis) vs. $log(t)$ (X-axis). The cross denotes
value (calculated by performing $F^\alpha$-integration numerically) to which 
straight line is fitted with slope 0.802 to extract 
the value of the exponent in equation (\ref{eq:L2} ), for a von Koch curve.}
\label{fig:secondm}
\end{figure}
In the case of von Koch curve $1/\alpha = 0.792$ and $1/2\alpha = 0.396$.
These expressions are in accordance with the numerical results presented in 
fig(\ref{fig:firstm}) and (\ref{fig:secondm}). In these figures a plot between
 moments and time is shown. These plots have been obtained by performing 
the $F^\alpha$ integrals in equations (\ref{eq:firstm}) and (\ref{eq:secondm})
numerically, for the von-Koch curve. From the plots one obtains
\[<L> \sim t^{0.403}\]
and 
\[<L^2> \sim t^{0.802}\]

Here we see a beautiful match between the expected values and numerical 
calculations within numerical accuracy. The plots also comfirm the 
expected  behaviour of $S_F^\alpha(u)$.
\section{Broad Distributions} \label{sec:bdist}

We have derived in section (\ref{sec:discreet}) , the expression for 
probability distribution for a discrete, simple random walk on fractal curve.
 Now we examine random 
walks with some different (non Gaussian) probability distributions for the 
individual steps.
\subsection*{Review of Broad Distributions} 
We begin by summarizing some relevant results for Broad distributions from
 \cite{Bouchaud}.

For a 1-D random walk in ordinary space, let $X_N$ denote
 the sum of  independent random variables $l_n$, i.e
\be
X_N = \sum_{n=1}^N l_n
\ee
where $l_n$ is the $n^{th}$ step size. 
Let us denote $p(l)$, the probability distribution for 
$l$(individual steps in the random walk) which is broad, that is ,it decreases 
for large $l$ as $l^{-1+\mu}$ with $\mu >0$. We consider two cases, 
\begin{itemize}
\item For  $0<\mu <1$ , $X_N$ , the first moment $\langle l \rangle$ is 
infinite.
\item For $1< \mu < 2$ , the first moment $\langle l \rangle $ is finite and 
the second moment $\langle l^2 \rangle $ is still infinite.
\end{itemize}
It is well known that limit distributions of the sum $X_N$ are
 defined by their characteristic functions which are given by $\tilde{L}_{
\beta,\mu}(k)$ in Fourier space, i.e $k$- space, when $\mu$ and $\beta$ are
parameters, $\beta$ being  the degree of asymmetry. We consider the case
$\beta =0$ for the stable distribution $L_{0,\mu}$. In this case large
 positive and negative values of $l_n$ occur with equal frequency. Then
\be  \label{eq:char}
\tilde{L}_{0,\mu}(k) = e^{-|k|^\mu}
\ee
Its Fourier transform  being
\be \label{eq:stable}
L_{0,\mu}(Z) = \frac{1}{2 \pi} \int_\infty^\infty dk e^{ikZ-|k|^\mu}
\ee
We set $Z = Z_N $ where,$ Z_N = X_N/N^{1/\mu}$ or $ (X_N-<l>N)
/N^{1/\mu}$ as $N \rightarrow \infty$ ,for $ 0<\mu<1$ or $1<\mu<2$ respectively. All the important results for such a case are given in \cite{Bouchaud}.
\subsection*{Broad Distributions on fractal curves}
Now we consider  an analogous approach for the random walk on fractal curve.
In particular we take  parametrizable fractal curve, the von-Koch curve
with $\gamma$- dimension equal to $log 4/log3$  as described in \cite{Abhay}.

We consider the following probability
 distribution in the limiting case, when $N \rightarrow \infty $.
We now abuse the above notation for obvious reasons to denote $L_{0,\mu}(\psi)
$ as the distribution in Fourier space of fractal curve $F$, where $\psi
= \bw(k)$ and $\Su(k) = J(\psi)$ as described in appendix.
Then 
\be
L_{0,\mu}(\psi) = \exp(-| J(\psi)|^\mu)
\ee
In what follows we will use $\tilde{L}$ to refer to quantity on the fractal
curve (space) and $\bar{L}$ to refer to the quantity in the real line.

The inverse Fourier transform (as defined in the appendix), 
for the above distribution is given by
\be
\tilde{L}_{0,\mu}(\theta)  = \frac{1}{2 \pi} \int_{C(-\infty,\infty)} 
\exp(i J(\psi)J(\theta)   -|J(\psi)|^\mu) d_F^\alpha \psi
\ee
The above integral in view of conjugacy between $F^\alpha$ integral and 
Riemann integral becomes:
\be \label{eq:conjequation}
 \bar{L}_{0,\mu} (y= J(\theta)) = \frac{1}{2 \pi}\int_\infty^{-\infty}
 \exp(i \tilde{k}y -|\tilde{k}|^\mu) d \tilde{k}
\ee
For $\mu = 2$ we obtain the gaussian distribution.
case:
\be
\bar{L}_{0,\mu}(y=J(\theta))= \frac{1}{\sqrt{2 \pi}} \exp(-\frac{ y^2}{2})
\ee
Applying the fractalizing operator to the above equation we get back the 
expression for distribution on fractal curve.
\be
\tilde{L}_{0,\mu}(\theta)= \frac{1}{\sqrt{2 \pi}} 
\exp(-\frac{\{J(\theta) = \Su(u)\}^2}{2})
\ee
Similarly for the Cauchy case when $\mu =1$ we get
\[\tilde{L}_{0,\mu} = \frac{1}{\pi(1+ \{J(\theta)=\Su(u)\}^2)}\]
Now, one can then discretize the integral in equation (\ref{eq:conjequation})
 by replacing $d\tilde{k}$ in terms of equal steps of size $\Delta \tilde{k}$,
 as follows:
\[\bar{L}_{0,\mu} (y= J(\theta)) = \sum_{m= -\infty}^{\infty} \exp[i 
\tilde{k}_m y - |\tilde{k}_m|^\mu] \Delta \tilde{k} \]
where $\tilde{k}_m = m \Delta\tilde{k}$.

For values of $\mu$ other than $1$ and $2$ , the expansion for large arguments
 is given by  \cite{Bouchaud}
\[\bar{L}_{\mu,0} (y = J(\theta)) = (\pi)^{-1} \sum_{m=1}^\infty (-)^{m+1}
\frac{y^{-(\mu m +1)}}{m!} \Gamma(1+m \mu) \sin(\pi \mu m / 2)\] 
The leading term of which is
\[\bar{L}_{\mu,0} (y= J(\theta)) = (\pi)^{-1} y^{-(\mu+1)}\Gamma(1+\mu) \sin
(\pi \mu/2)\]
Applying the fractalizing transformation to the above we get the
 result:
\begin{eqnarray} \label{eq:result}
\tilde{L}_{\mu,0} (\theta) &=& (\pi)^{-1} \Su(u)^{-(\mu+1)}\Gamma(1+\mu) \sin
(\pi \mu/2)\\
& \sim & L^{-\alpha(\mu+1)} 
\end{eqnarray}
for large values of $J(\theta) = \Su(u)$ or $L$ where $L$ denotes the
 Euclidean distance.

One can see that 
$<L>$ is infinite for $\mu \leq 1/\alpha$ and is finite for $\mu > 1/\alpha$
for these large values of $L$. Also the second moment $<L^2>$ diverges for 
$\mu \leq 2/\alpha$ and is finite for $\mu > 2/\alpha$ for large values of $L$.
 Hence we see a scaling 
down of the absolute moments for Levy distrubutions due to the fractal 
structure of the underlying space.
\section{First Passage Time} \label{sec:fpt}
The first passage time is a standard problem in Statistical Physics 
\cite{Feller,Grimmet}. 
Here we extend it to first passage time on fractal curves.

Random walk on a fractal curve can be described by forward and 
backward steps along the curve that cover equal mass on the fractal curve
as in section (\ref{sec:discreet}), so
 that there is one- to one correspondence between random walk on a straight 
line (the parameter u) and that on the curve $F$.

The first passage time, or the time required to reach a certain point 
$\theta = \mathbf{w}(u)$ on the curve $F$, for the first time can be defined
as follows:
\begin{definition}
The first passage time $T(\theta)$ to a point $\theta \in F$ is given by
\be
T(\theta) = \inf\{ t : \Theta(t) = \theta\}
\ee
where $\Theta(t) $ is the position of a random walker on $F$ at time $t$, which
started at a point $\theta_0$ at time $t = 0$ .
\end{definition}
Now we explore the relation between this first passage time and the mass
 covered by the random walker on the fractal curve in this time.

When the walker performs forward and backward steps along the curve, let the
 mass covered in the $n^{th}$ step be given by 
$\gamma_F^\alpha(F,u_n,u_{n-1})$, corresponding to parameter value lying 
between $u_{n-1}$ and $u_n$.
 The resultant displacement ( considered on the $\Su$-axis), of the particle 
performing a random walk of $M$ steps, some of which may be in forward
direction and rest in the backward direction on the curve, so as to reach
an arbitrary point $\bw(u_k)$ on $F$,
 is given by sum of $\gamma_F^\alpha $ over $M$ steps. For $k < M$ such that
 $M = k+r$, and step
size $\Delta$, for every sequence $\{u_1,u_2, \dots, u_M\}$, such that, 

\be \label{eq:resultant}
S_F^\alpha(u_M) = k \Delta
\ee

Let the time duration between consecutive steps on the fractal curve $F$, be
 denoted by $\tau$. We calculate the maximum mass that a random walker covers 
in a given
time $t$, such that $t= M \tau$, where $M$ is the total number of steps taken
in time $t$.  

Since $k = M-r$ we can write equation (\ref{eq:resultant}) in the following 
form:
\be
S_F^\alpha(u_M) = (\frac{t}{\tau} -r) \Delta
\ee
It can be clearly seen that $S_F^\alpha(u_M)$ is maximum when $r =0$ on the
rhs of the above equation.
Hence
\be
S_F^\alpha(u_M)|_{max} = \frac{t}{\tau} \Delta
\ee

Next we calculate the minimal time required by a random walker to cover certain
mass on the curve.

Conversely, let $t = M \tau$ be the time taken for $M$ steps on the curve $F$,
 also let $M = k+r$ as given above, then  $ k = \frac{t-r\tau}{\tau}$, and 
we can write equation \ref{eq:resultant} as
\be 
S_F^\alpha(u_M) = \frac{ t - r \tau}{\tau} \Delta
\ee
Thus,
\[t = r \tau + \frac{\tau S_F^\alpha(u_M)}{\Delta} \]
Hence we see that $t$ is minimum when $r =0$, Hence
\be
t_{min} = \frac{\tau}{\Delta} S_F^\alpha(u_M)
\ee 
 
\begin{figure}[h!]
\resizebox{\columnwidth}{!}{\includegraphics{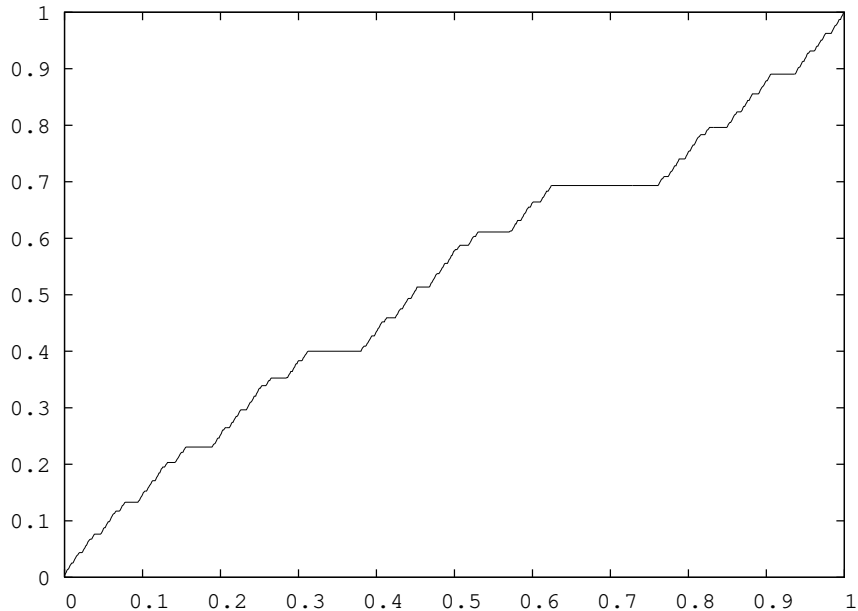}}
\caption{ Plot of $L_{max}$ (Y-axis) vs $t_{min}$ (X-axis) for random walk on
the von Koch curve, obtained numerically.}
\label{fig:lmax}
\end{figure}
An interesting quantity, which can be analysed is the maximum Euclidean
 distance on a fractal curve, covered by the random walker in a minimum
 time $t$.
 In general these will depend on the geometry of the particular
 fractal curve of interest. The relation between 
Euclidean distance on a fractal curve and the staircase function \cite{Abhay}
 can only be obtained numerically, even for a simple fractal curve like the
 von-Koch curve. 
A plot between $t_{min}$ minimum time required to reach a point and $L_{max}$ 
(maximum Euclidean distance) calculated numerically for a von Koch curve, is 
shown in fig (\ref{fig:lmax}). One may note a step-like behaviour in this 
figure. We comment on this in the results.
\section{Results}
It is clear from the figures (\ref{fig:firstm}) and (\ref{fig:secondm}) that
a subdiffusive behaviour is addressed by the gaussian-like random walks on the
 von Koch curves.

Eq. (\ref{eq:result}) shows the behaviour of the stable law $L_{\mu,0}$ for
 values of $\theta$ for which $\Su(u)$ is large. 
This 
behaviour differs from the ordinary law for large deviation, when the
underlying space is not a fractal and when the stable law decreases as 
$y^{-(1+\mu)}$. In the case of an underlying fractal space, the exponent  
gets multiplied by a factor of $\alpha$ the dimension of the fractal curve 
itself. 
Thus we see that from a Euclidean perspective, the behaviour gets a direct 
contribution from the exponent of the distribution and the dimension of 
the curves. This is a striking difference. When taking 
straight line approximations on fractal curves, the
dimension of the fractal curve contributes only indirectly. While using the
 above 
method, the fractal dimension plays a direct role in changing the nature 
of heavy tails of the distribution. 

In fig. (\ref{fig:lmax}) we see a step-like behaviour, this explicitly
shows that the Euclidean distance on the curve remains constant for a short
 while and then rises. There is a largest period of constant Euclidean 
distance in the figure, this depends on the geometry and the initial position 
of the random walker on the von Koch curve. Here we calculate the 
Euclidean distance from the origin or initial point (0,0) on the curve 
in $\mathbf{R}^2$ .
\section*{Appendix 1}

\textbf{The Fourier Transform}

From the definition of conjugacy \cite{Abhay}
\be \label{eq:defcon}
\phi[f] (\Su(u)) = f(\mathbf{w}(u))
\ee
The  Fourier Transform on the real line, for a function $g(v)$, is defined
 by
\be
g(v) = \int_{-\infty}^{\infty} \tilde{g}(y) \exp(-ivy) dy 
\ee
and the inverse Fourier Transform is
 
\be \label{eq:invtrans}
\tilde{g}(y) = \frac{1}{2 \pi}\int_{-\infty}^{\infty} g(v) \exp(ivy) dy 
\ee
In the case of a parametrizable fractal curve $F$, which is obtained
by a fractalizing transformation on an interval of the real line, we propose
 that the Fourier space can also be obtained by the same fractalizing 
transformation on the interval of a real line. The interval may be $(-\infty,
\infty)$.

 Let $\tilde{g} = \phi[\tf]$ and $g = \phi[f]$ also $v=\Su(k)$ and $y
= \Su(u)$.

We use the notation $J(\theta) = \Su(u)$ and $J(\psi) = \Su(k)$, where
$\theta = \bw(u)$ and $\psi = \bw(k)$.

Then taking Fourier transform of the LHS of equation (\ref{eq:defcon}) one can
 write
\begin{eqnarray} \label{eq:fourier}
\tilde{\phi[f]} (v=J(\psi)) &=& \int_{-\infty}^\infty \phi[f](y =J(\theta))
\exp(-iyv)dy \nonumber \\
& = &  \int_{C(-\infty,\infty)} f(\theta) \exp(-iJ(\theta) v)d_F^\alpha \theta 
\end{eqnarray}
 where 
\[C_(-\infty,\infty) = \lim_{a \rightarrow -\infty, b \rightarrow \infty} 
C_(a,b) \]
and 
\begin{eqnarray} \label{eq:fourier1}
\phi[\tf] (v=J(\psi)) & = & \phi [ \int_{-\infty}^\infty f(y =J(\theta)) 
\exp(-iyv)dy] \nonumber \\
& = &  \int_{C(-\infty,\infty)}f(\theta) \exp(-iJ(\theta) v)d_F^\alpha \theta
\end{eqnarray}
Comparing equations (\ref{eq:fourier}) and (\ref{eq:fourier1}), we can write
\[\tilde{\phi[f]} = \phi [\tilde{f}] \]
Also one can define the action of $\phi$ in Fourier space as
\be
\phi[\tilde{f}](v = J(\psi)) = \tf(\psi)
\ee
Now using conjugacy one can rewrite equation (\ref{eq:fourier}) as 

\be
\tf(\psi) =  \int_{C(-\infty,\infty)}f(\theta)
\exp(-i J(\theta) J(\psi) ) d_F^\alpha \theta
 \ee
Similarly, inverse transform of the above can be obtained from 
equation (\ref{eq:invtrans}), which can be 
written as

\be
f(\theta) = \frac{1}{2 \pi} \int_{C(-\infty,\infty)}\tf(\psi)
\exp(i J(\theta) J(\psi) ) d_F^\alpha \theta
 \ee
\section*{Appendix 2}
\textbf{Review of Calculus on Fractal Curves}

 For a set $F$ and a subdivision $P_{[a,b]}, a<b$, $[a,b]
\subset [a_0,b_0]$ let $\mathbf{w}:[a,b] \rightarrow F$, then we define the 
mass function as follows:

\be \gamma^{\alpha}(F,a,b) = \lim_{\delta \rightarrow 0} \inf_{P_{[a,b]}:|P|
 \leq \delta} \sum_{i=0}^{n-1} 
 \frac{|\mathbf{w}(t_{i+1}) - 
\mathbf{w}(t_i)|^\alpha}
{\Gamma(\alpha +1)}  \label{eq:sigma}\ee
where $|\cdot|$ denotes the euclidean norm on $\mathbf{R^n}$, $1 \leq \alpha
\leq n$ and $P_{[a,b]} = 
\{a=t_0,\ldots,t_n=b\}$. 

The staircase function, which gives the mass of the curve upto a certain point
on the fractal curve $F$ is defined as
\be
S_F^{\alpha}(u) = \left\{ \begin{array}{ll}
	\gamma^{\alpha}(F,p_0,u) & u \geq p_0 \\
	- \gamma^\alpha(F,u,p_0) & u < p_0
		\end{array}
	\right. 
	\label{eq:staircase_function} \ee
where $u \in [a_0,b_0]$.

A point on the curve $\mathbf{w}(u) \equiv \theta $ and $S_F^\alpha(u) \equiv
J(\theta)$.
The $F^\alpha$ derivative is defined as:

\be (D_F^\alpha f)(\theta)= F \mbox{-}\lim_{\theta' \rightarrow \theta} 
\frac{f(\theta')-f(\theta)}
{J(\theta')-J(\theta)} \label{eq:derivative}\ee
 
The $F^\alpha$-integral is also defined and is denoted by
\be \int_{C(a,b)} f(\theta) d_F^\alpha \theta
\ee 
where $C(a,b)$ is the section of the curve lying between points $\mathbf{w}(a)$
 and $\mathbf{w}(b)$ on the fractal curve $F$. 
$\bullet$

\end{document}